# Model Confidence-Guided Multi-Image Fusion of Fundus Images for Diabetic Retinopathy Diagnosis

Model Confidence-Guided Multi-Image Fusion

***Ananya Raghu[1*], Anisha Raghu[1*], Alice S. Tang[4], Yannis M. Paulus[2,3], Tyson N. Kim[5**†], Tomiko T. Oskotsky[4,6**†]***

*[1]Massachusetts Institute of Technology, Cambridge, MA, USA*

*[2]Wilmer Eye Institute, Department of Ophthalmology, Johns Hopkins University, Baltimore, MD, USA*

*[3]Department of Biomedical Engineering, Johns Hopkins University, Baltimore, MD, USA*

*[4]Bakar Computational Health Sciences Institute, University of California San Francisco, San Francisco, CA, USA*

*[5]Department of Ophthalmology, University of California San Francisco, San Francisco, CA, USA*

*[6]Division of Clinical Informatics and Digital Transformation, University of California San Francisco, San Francisco, CA, USA*

****Co-first authors: these authors contributed equally to this work.***

***[**]Co-senior authors: these authors jointly supervised this work.***

***[†]Corresponding Authors Email Addresses:***

Tomiko.Oskotsky@ucsf.edu, Tyson.Kim@ucsf.edu

## Structured Abstract

*Purpose*

Early screening for eye diseases is critical, in low- and middle-income countries where access to care is limited. We investigate whether a confidence-guided, multi-image diabetic retinopathy diagnosis framework can integrate image filtering with confidence-aware predictions to enable reliable screening at image capture.

*Methods*

We develop a multi-image fusion method that aggregates retinal views to improve confidence and balanced accuracy. Our method also uses confidence to identify unreliable predictions, prompting retake when needed.

We compare: (1) a cascaded pipeline consisting of an image quality model followed by a disease diagnosis model using a single image per patient, (2) confidence-based prediction, and (3) our confidence-based multi-image fusion pipeline. All methods are evaluated using a RETFoundGreen backbone on the mBRSET (n = 1,234 patients, mobile captured) and BRSET (n = 7,599 patients, tabletop clinic captured) datasets.

*Results*

At 70% coverage, our method achieves 91% balanced accuracy on mBRSET and 97% on BRSET, improvements of ~12% and ~6%, respectively, over cascade filtering. The image-quality cascade reaches a sensitivity of 61% on mBRSET and 86% on BRSET, whereas our framework reaches higher sensitivities of 94% and 96%, respectively, at 50% coverage.

*Conclusions*

Human-annotated quality labels are weakly associated with diagnostic performance, and confidence-based filtering consistently outperforms image quality-based cascaded pipelines.

*Translational Relevance*

*Using integrated confidence-based multi-image fusion, patients receive more reliable predictions, reducing incorrect diagnoses during screening. Because the framework uses a lightweight backbone and requires a single inference pass per image, it may be compatible with low-latency mobile screening systems in resource-limited settings.*

## Introduction

Diabetic retinopathy (DR) is one of the leading causes of blindness in the world and is rapidly increasing in prevalence.[1] Early screening is essential, yet access to medical care can be limited in low and middle income countries (LMICs). There is a significant shortage of doctors in LMICs, which contain over 90 percent of the world's blind population.[2]

DR is a complication of diabetes caused by chronic high blood sugar levels, which leads to damage of blood vessels in the eye. This can cause diabetic macular edema (DME), a condition where fluid accumulates in the macula, causing blurred vision and significant vision loss if left untreated. Diabetic retinopathy, including vision-threatening complications such as diabetic macular edema, is a leading cause of vision loss and a growing global health concern, affecting over 103 million people worldwide.[3] Early detection and proper management can help slow its progression and reduce the risk of severe vision impairment, making medical screening, particularly accessible automated screening methods for medically underserved areas, very important.[4]

Recently, smartphone fundus imaging has emerged as a cost effective and accessible alternative for disease screenings, allowing for integration with rapid cloud-based workflows for diagnosis and enabling smartphone-based diagnosis of diseases.[5-10] Despite these advances, two key challenges remain.

The first challenge is assessing image quality. Several earlier approaches have explored fundus image quality assessment as a standalone task, focusing on classifying images as gradable or ungradable based on factors such as illumination, sharpness, positioning, or retinal structure visibility[11-13]. These methods range from hand crafted feature pipelines[11] to convolutional neural network (CNN)-based and transfer-learning approaches[12-13], and many report strong performance on image quality benchmarks. While prior work has developed sophisticated quality assessment systems, these approaches often evaluate image quality in isolation and provide only modest improvements when integrated into downstream diagnostic pipelines.

Second, leveraging multiple images acquired from each eye to produce a robust patient-level diagnosis remains an open challenge. An existing approach to multi-image aggregation is limited to complex binocular fusion methods that are constrained to one image per eye.[14] Although this approach can improve performance, it increases system complexity and limits deployability in low-resource environments.

To address these challenges, we propose a confidence-guided multi-image fusion pipeline for diabetic retinopathy diagnosis. Our method integrates:

- Patient level aggregation using transformer based architecture
- A second stage confidence threshold to ensure reliable final predictions.

Unlike prior approaches, our method uses model confidence as an implicit indicator of image quality, eliminating the need for explicit quality labels and extrinsic workflows. Additionally, it is a lightweight, fully offline pipeline suitable for mobile deployment in resource-limited settings.

## Methods

*Datasets*

In this work, we primarily use the Mobile Brazilian Multilabel Ophthalmological Dataset (mBRSET)[15] dataset, which contains 5,164 images captured on smartphone-based retinal fundus cameras from 1,291 patients, containing labels for diabetic retinopathy. We also make use of the Brazilian Multilabel Ophthalmological Dataset (BRSET).[16] This dataset contains over 16,266 tabletop clinic-captured fundus images from 8,524 Brazilian patients, containing disease diagnosis information for diabetic retinopathy.

*Data Preprocessing*

Diabetic retinopathy labels were binarized from the original ICDR (International Clinical Diabetic Retinopathy) severity scale[17], with non-zero grades mapped to DR-positive and grade 0 mapped to DR-negative. For mBRSET, missing ICDR labels were imputed using the corresponding paired image from the same eye. For BRSET, an image was labeled as poor quality if any evaluated feature, including focus, illumination, image field, or artifacts, was graded as abnormal, whereas for mBRSET, a single overall image quality label was directly provided based on prior human assessment.

To support our multi-image fusion experiments, we restricted the datasets to patients with a fixed number of images per eye: one image per eye for BRSET and two images per eye for mBRSET. Patients with missing or incomplete image sets were excluded. We did not require diabetic retinopathy labels to be consistent across both eyes, since DR can present asymmetrically and may be present in only one eye.[18] Additionally, because model performance was evaluated at the patient level, each patient was assigned a single diabetic retinopathy label by taking the maximum DR label across all included eye-level images for that patient.[19-20]

Additional details on preprocessing steps are provided in Supplementary Tables S1 and S2. The class distribution of image quality and diabetic retinopathy labels for both datasets is summarized in Tables 1 and 2.

After preprocessing, the final BRSET cohort contained 15,198 images and the mBRSET cohort contained 4,936 images. Both datasets were split into training, validation, and test sets using a 70%, 15%, 15% split at the patient level to prevent data leakage (Supplementary Table S3 and S4).

| Variable | Category | N (%) |
|---|---|---|
| Image Quality | Good | 4767 (96.6%) |
| | Bad | 169 (3.4%) |
| Diabetic Retinopathy | Present | 1262 (25.6%) |
| | Absent | 3674 (74.4%) |

***Table 1:*** *mBRSET data distribution*

| Variable | Category | N (%) |
|---|---|---|
| Image Quality | Good | 13380 (88.0%) |
| | Bad | 1818 (12.0%) |

| Diabetic Retinopathy | Present | 973 (6.4%) |
| --- | --- | --- |
| | Absent | 14225 (93.6%) |

***Table 2:*** *BRSET data distribution*

*Single-Image Based Diabetic Retinopathy Detection Strategies*

Our first objective is to determine whether image quality can serve as a reliable predictor of a model's diagnostic accuracy. To investigate this, we compare two complementary image-level filtering strategies.

The first baseline is a cascaded approach consisting of an image quality assessment model followed by a disease diagnosis model (Figure 1a). In this setup, image quality predictions from a dedicated quality model are used to filter input images before disease prediction. Specifically, the output probabilities from the image quality model are compared against a range of thresholds, allowing the strictness of filtering to be varied. Images predicted to be below the selected quality threshold are excluded and the remaining images are passed to the diagnosis model.

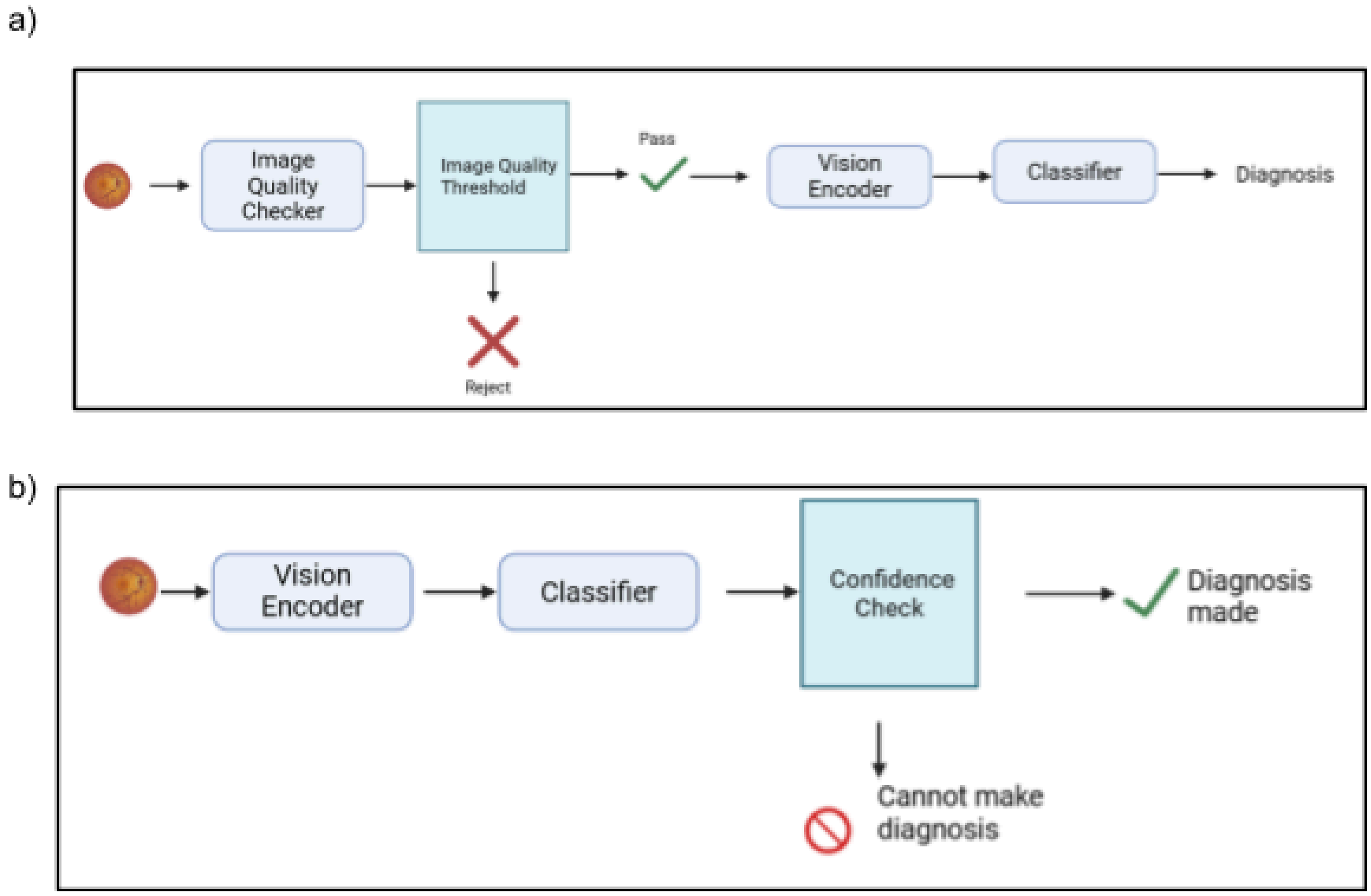

***Figure 1:*** *Comparison of image filtering strategies: (a) a cascaded pipeline that removes low-quality images prior to diagnosis using explicit quality metrics, and (b) a confidence-based approach that rejects images based on the diagnostic model's prediction confidence.*

The second approach for single image based diabetic retinopathy detection (Figure 1b) is a model confidence-based rejection strategy that does not rely on explicit quality annotations. Instead, we apply confidence-based filtering by sweeping over probability thresholds and retaining only test images whose predictions fall within the specified confidence bounds. This allows the system to use the model's own uncertainty as an implicit indicator of image quality.

*Backbones and training*

All image encoders, including both the diabetic retinopathy diagnosis models and the image quality assessment model used in the cascaded baseline, were initialized with RETFoundGreen weights, a lightweight retinal foundation model that provides generalizable retinal representations. Our model is built on RETFoundGreen, which is approximately 14× smaller than server-side retinal foundation models such as RETFound, enabling efficient on-device inference on mobile devices where deployment of larger models is impractical.[21-22] The image quality assessment model used a binary output head trained to predict image quality, while the diabetic retinopathy models used a binary output head trained to predict diabetic retinopathy. The backbone follows a Vision Transformer small architecture[23], where each input image is divided into fixed-size 16 × 16 patches, enabling the model to capture contextual relationships across the fundus image.

All models were trained using a weighted cross-entropy loss to account for class imbalance, as shown in Equation 1, where $N$ is the number of samples in the training set, $w_0$ is $\frac{N_1}{N}$ and $w_1$ is $\frac{N_0}{N}$, where $N_0$ and $N_1$ are the number of samples of class 0 and class 1 in the dataset, $y_i$ and $p_i$ are the true label and predicted probability for sample $i$ respectively.

$$L = -\frac{1}{N}\sum_{i=1}^{N}[w_1 y_i \, log(p_i) + w_0 (1 - y_i) \, log(1 - p_i)] \qquad (1)$$

Optimization was performed using Adam[24], and models were trained for up to 20 epochs with early stopping based on validation performance, using a patience of 3 epochs. All fundus images were resized to a resolution of 392 × 392 pixels prior to training. We applied image augmentation techniques including rotation, flipping, and cropping during training.

*Multi-Image Fusion Strategies*

Our next objective is to evaluate the impact of leveraging multiple fundus images for diabetic retinopathy detection. To this end, we compare simple late fusion techniques: mean and max based pooling with transformer-based fusion approaches. For mean and max fusion, we use a shared RetFoundGreen model backbone that outputs probabilities per image. The probabilities are pooled with either mean or max pooling to produce the patient level probability. For the transformer fusion model, every image is converted to an embedding through a shared image backbone, followed by two transformer encoder layers to produce the patient-level prediction. The backbone and the transformer encoder layers are trained end-to-end. For mBRSET, a single four-image transformer model was trained using all available images per patient.

To enable fair comparison across single-image and multi-image approaches, all methods were evaluated at the patient level using the patient-level diabetic retinopathy label described above. For single-image approaches, we evaluated each possible image position independently: two images per patient for BRSET and four images per patient for mBRSET. Performance was then averaged across these image choices, representing a setting in which only one representative fundus image is available per patient.

For two-image approaches, BRSET used both available images, corresponding to one image from each eye. For mBRSET, where each patient had two images per eye, we evaluated all four possible left/right image pairings by selecting one left-eye image and one right-eye image. To evaluate two-image inference without training an additional model, we used the trained four-image mBRSET transformer and masked the embeddings for the two omitted images. Performance was then averaged across the four left/right pairings. For four-image approaches on mBRSET, all four images for each patient were used simultaneously. All experiments were repeated across multiple trained model checkpoints, and reported values are presented as mean ± standard deviation across runs.

*Proposed Framework*

We integrate the ideas of multi-image fusion and model confidence based rejection of patient images in our approach. Our method is designed to improve the reliability of diabetic retinopathy detection by leveraging multiple images per patient (Figure 2). A confidence check is applied at the patient level, retaining only those patients whose aggregated probability is strongly confident toward either class. For patients who pass this check, the mean probability is used to generate the final diagnosis.

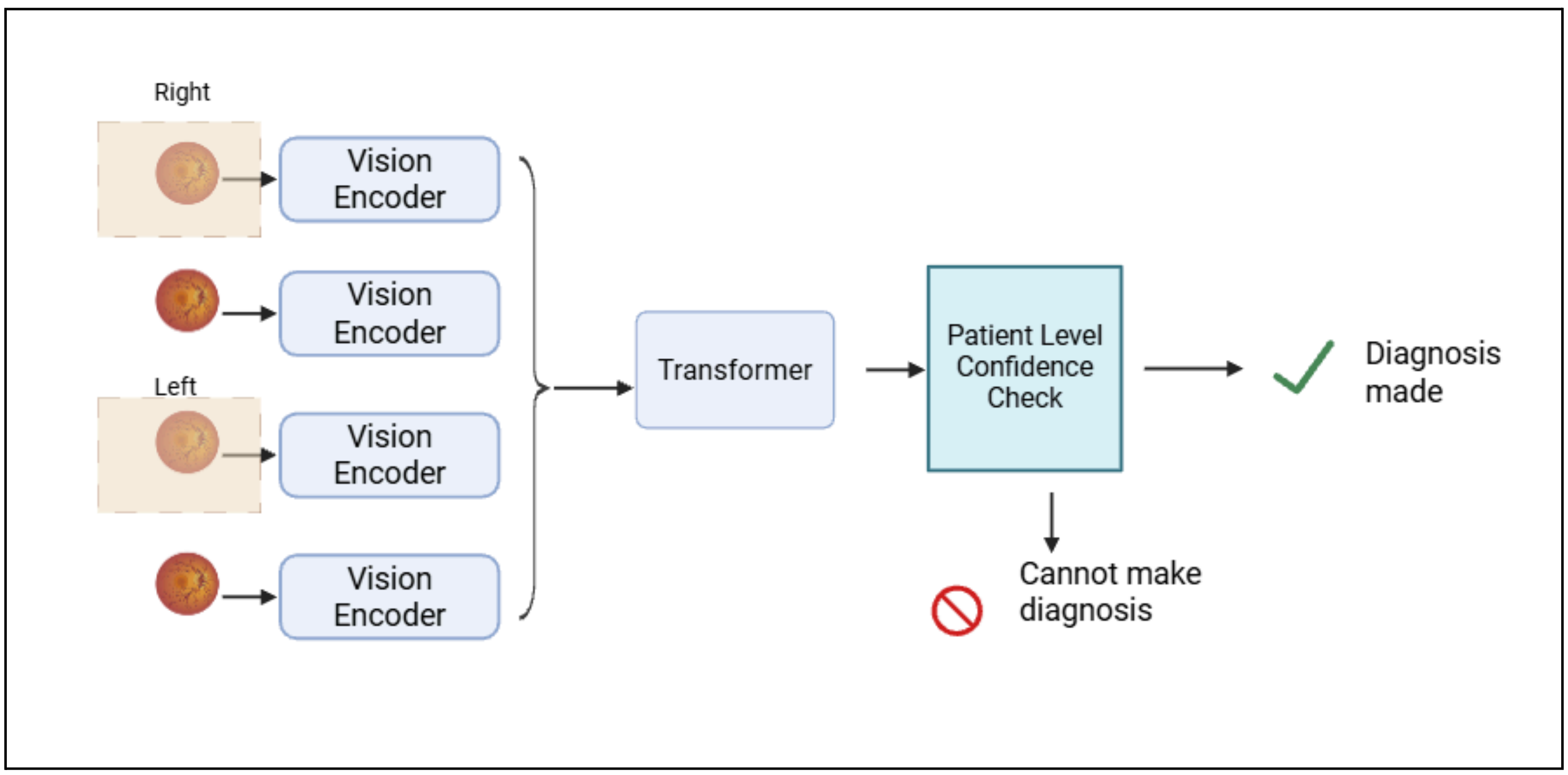

***Figure 2:*** *Confidence-guided multi-image fusion pipeline. Fundus images from both eyes are processed with a shared vision encoder, and the resulting embeddings are fused using a transformer to generate a patient-level diabetic retinopathy prediction. The transformer supports a number of input images through masking, allowing the same architecture to be evaluated with different image combinations.*

We parameterize confidence-based filtering using symmetric margins around 0.5 rather than independent lower and upper thresholds. Specifically, a threshold margin of $T$ defines the acceptance region

$p < 0.5 - T$ or $p > 0.5 + T$ where $p$ denotes the model's predicted probability.

We compute the balanced accuracy, sensitivity, and coverage for each threshold choice on the test set.

*Evaluation Metrics*

Model performance was evaluated using a set of metrics. Specifically, we report balanced accuracy (BA), sensitivity (recall), specificity, area under the receiver operating characteristic curve (AUROC), area under the precision-recall curve (AUPRC), F1 score, and positive predictive value (PPV).

## Results

We first evaluate the single-image baseline and patient-level fusion strategies on the mBRSET and BRSET datasets (Tables 3 and 4). Fusion-based approaches consistently outperform the single-image baseline, highlighting the benefit of leveraging information from multiple fundus images when generating patient-level predictions. Among the evaluated methods, transformer-based fusion achieves the strongest overall performance across both datasets. It also exhibits lower variability, as reflected by its smaller standard deviations, suggesting more stable performance than the other fusion approaches.

| Model | BA | Sensitivity | Specificity | AUPRC | AUROC | F1 | PPV |
|---|---|---|---|---|---|---|---|
| Single Image Baseline | 0.77 ± 0.04 | 0.70 ± 0.11 | 0.84 ± 0.16 | 0.81 ± 0.02 | 0.85 ± 0.02 | 0.69 ± 0.05 | 0.73 ± 0.16 |
| Mean Fusion | 0.80 ± 0.05 | 0.75 ± 0.13 | 0.85 ± 0.21 | 0.88 ± 0.01 | 0.91 ± 0.01 | 0.74 ± 0.07 | **0.80 ± 0.20** |
| Max Fusion | 0.76 ± 0.09 | **0.88 ± 0.08** | 0.64 ± 0.25 | 0.89 ± 0.01 | 0.91 ± 0.01 | 0.68 ± 0.10 | 0.58 ± 0.14 |
| Four Image Transformer-based Fusion | **0.84 ± 0.01** | 0.79 ± 0.02 | **0.89 ± 0.02** | **0.90 ± 0.01** | **0.92 ± 0.01** | **0.78 ± 0.01** | 0.77 ± 0.04 |

***Table 3:*** *Comparison of single-image baseline and patient-level fusion strategies on the mobile mBRSET dataset. Bold/highlighted values indicate the highest mean performance per metric.*

| Model | BA | Sensitivity | Specificity | AUPRC | AUROC | F1 | PPV |
|---|---|---|---|---|---|---|---|
| Single Image Baseline | 0.90 ± 0.01 | 0.85 ± 0.03 | 0.95 ± 0.02 | 0.84 ± 0.01 | 0.94 ± 0.01 | 0.71 ± 0.05 | 0.60 ± 0.08 |
| Mean Fusion | 0.91 ± 0.01 | 0.86 ± 0.02 | 0.95 ± 0.01 | 0.89 ± 0.01 | 0.96 ± 0.01 | **0.78 ± 0.06** | **0.72 ± 0.09** |
| Max Fusion | 0.91 ± 0.02 | **0.92 ± 0.02** | 0.92 ± 0.02 | 0.89 ± 0.01 | 0.96 ± 0.01 | 0.64 ± 0.07 | 0.50 ± 0.08 |
| Two Image Transformer-based Fusion | **0.92 ± 0.01** | 0.87 ± 0.04 | **0.96 ± 0.02** | **0.90 ± 0.02** | **0.97 ± 0.01** | 0.73 ± 0.05 | 0.65 ± 0.10 |

***Table 4:*** *Comparison of single-image baseline and patient-level fusion strategies on the tabletop clinic BRSET dataset. Bold/highlighted values indicate the highest mean performance per metric.*

We next evaluate how different fusion strategies trade off diagnostic performance against patient coverage under selective acceptance on both mBRSET and BRSET datasets. Compared to cascaded image quality filtering, confidence-based rejection achieves higher balanced accuracy across coverage levels (Figure 3 and 4). Multi-image fusion further improves performance, with two- and four-image models maintaining higher balanced accuracy than both single-image and cascade approaches at matched coverage levels.

a)

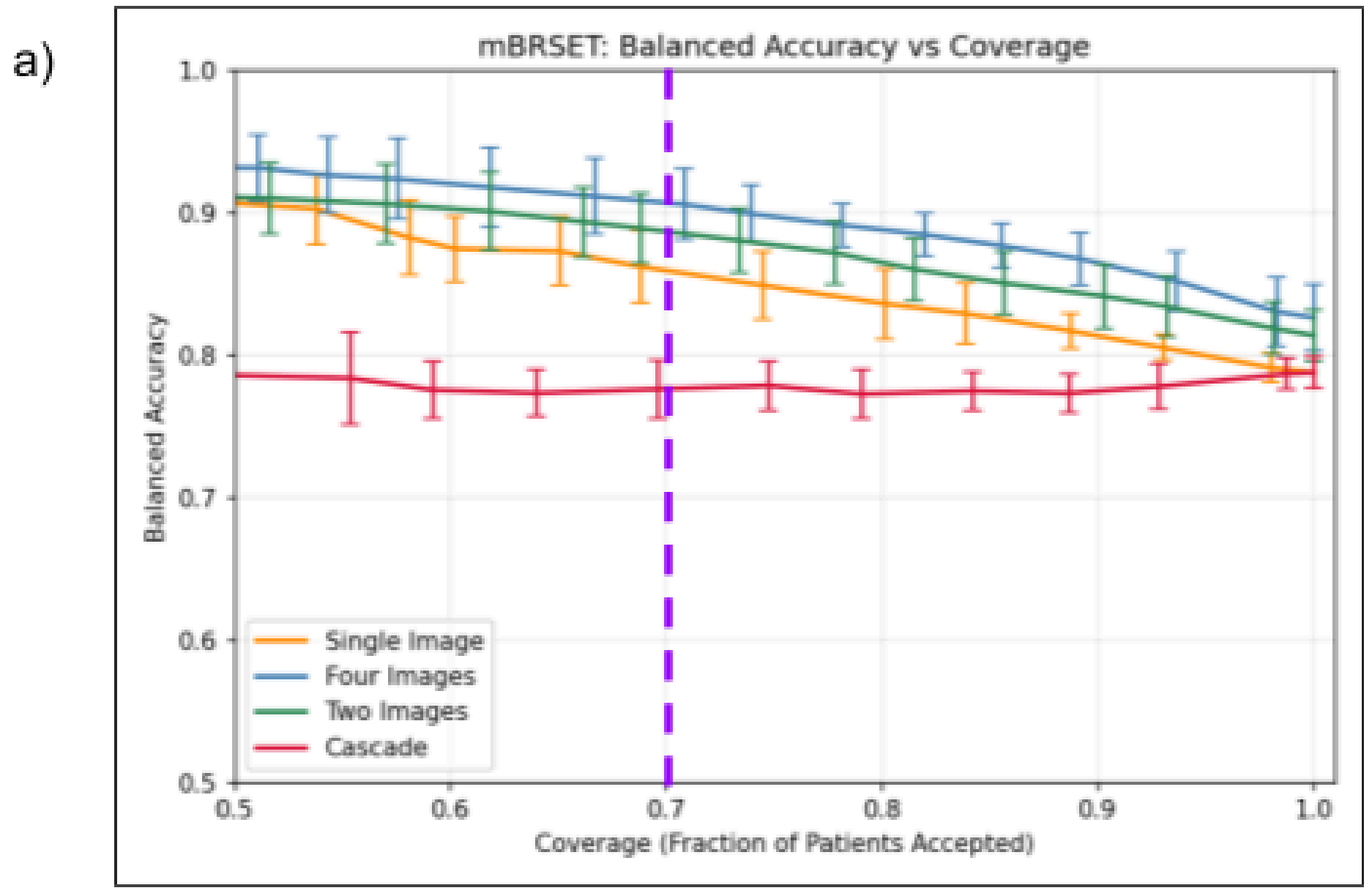

b)

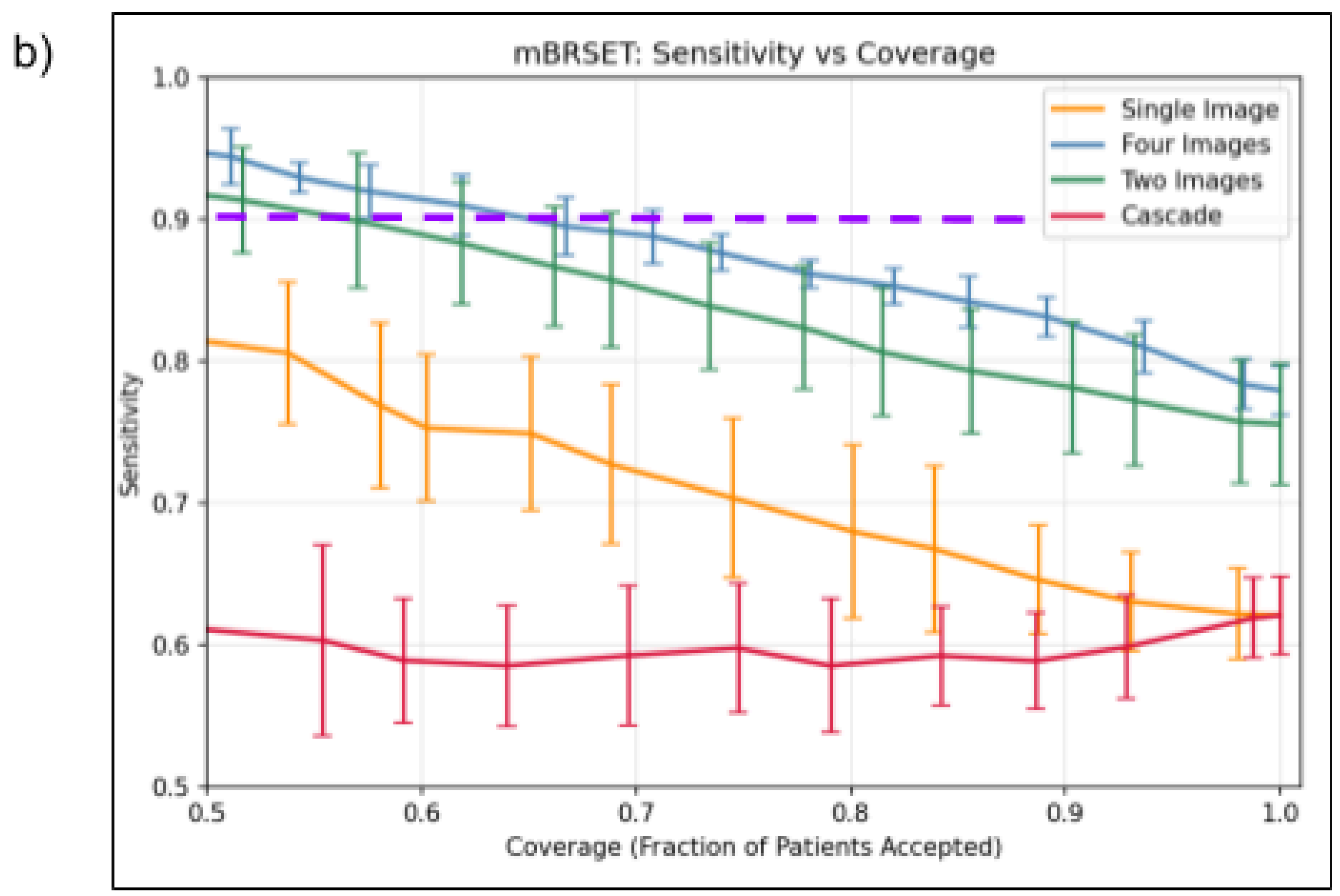

***Figure 3:*** *The impact of confidence based filtering vs image quality based filtering and the benefits of having multiple images per patient are shown for (a) Balanced accuracy and (b) sensitivity on the mBRSET dataset. Rejecting low-confidence patients generally improves performance, with two- and four-image fusion outperforming single-image and cascade baselines at matched coverage. The purple dashed line denotes 70% patient coverage in panel (a) and the 90% sensitivity operating point in panel (b).*

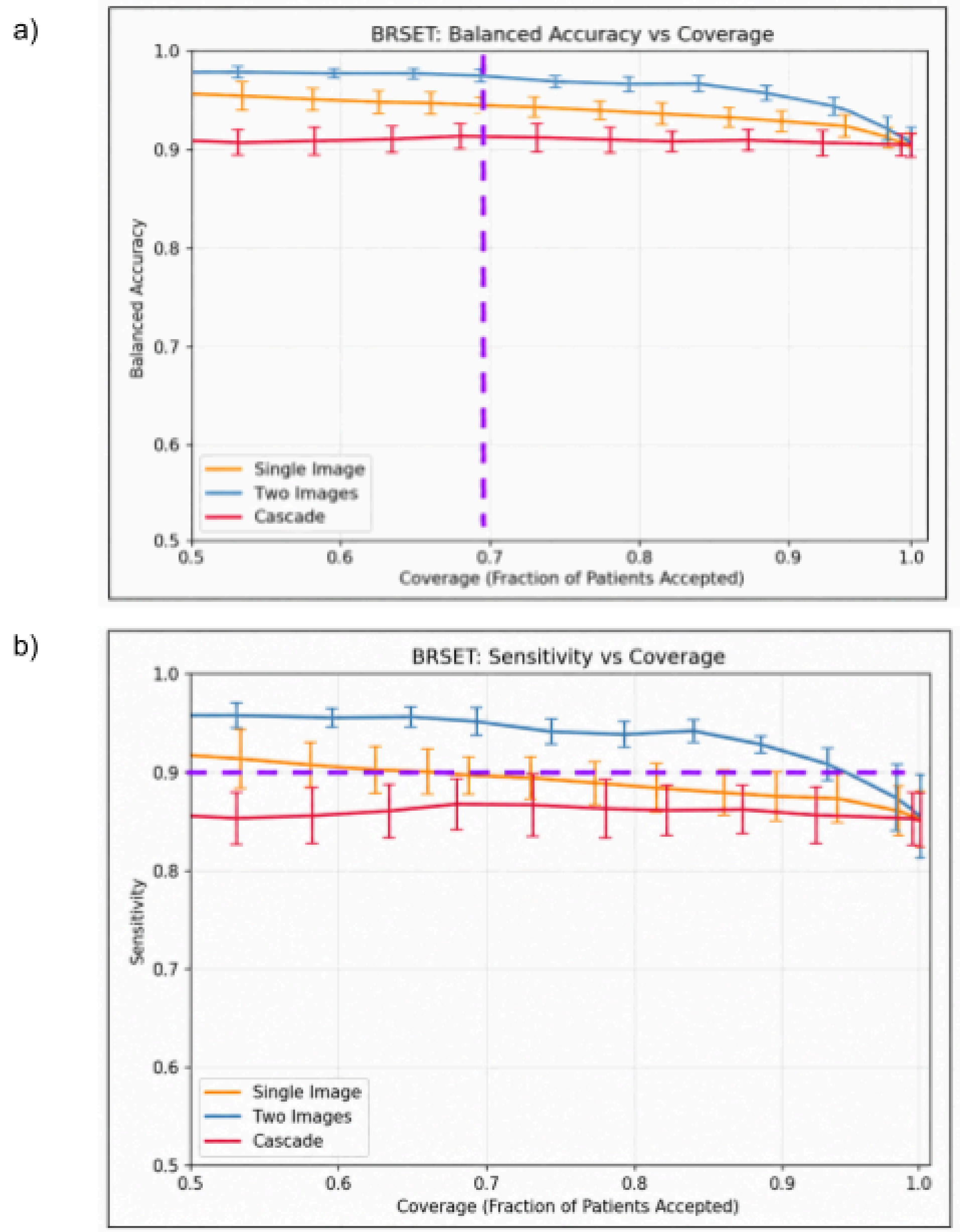

***Figure 4:*** *The impact of confidence based filtering vs image quality based filtering and the benefits of having multiple images per patient are shown for (a) Balanced accuracy and (b) sensitivity on the BRSET dataset. The purple dashed line denotes 70% patient coverage in panel (a) and the 90% sensitivity operating point in panel (b).*

On the mBRSET dataset, Figure 3a shows that at 70% patient coverage, the image-quality cascade achieves approximately 78% balanced accuracy, whereas confidence-based single-image prediction reaches approximately 86%, and confidence-based multi-image fusion reaches approximately 90-91%, with the four-image fusion model performing highest. Figure 3b further highlights the sensitivity-coverage tradeoff under confidence-based deferral: at the 90% sensitivity operating point, four-image fusion retains a larger fraction of patients than the two-image fusion, with an approximate 10-15% percentage-point gain in coverage.

On the BRSET dataset, Figure 4a shows a similar trend: at 70% patient coverage, our confidence-based two-image fusion achieves approximately 97% balanced accuracy, compared with approximately 91% for cascade filtering and 94-95% for the single-image approach. This corresponds to an approximately 6 percentage-point improvement over cascade filtering and a 2-3 percentage-point improvement over the single-image approach. Figure 4b further shows that at the 90% sensitivity operating point, our two-image fusion retains approximately 90-95% patient coverage, compared with approximately 65-70% for the single-image approach, corresponding to an approximately 20-25 percentage-point gain in coverage. Together, these results highlight the robustness of the proposed pipeline across both the mBRSET and BRSET datasets, demonstrating improved coverage while maintaining strong balanced accuracy and sensitivity.

To determine whether the weaker performance of the cascaded pipeline was due to poor image-quality classification, we separately evaluated the standalone image quality assessment model on the test set.

| **Model** | **BA** | **Sensitivity** | **Specificity** | **AUPRC** | **AUROC** | **F1** | **PPV** |
|---|---|---|---|---|---|---|---|
| mBRSET Image Quality Checker | 0.75 | 0.97 | 0.58 | 0.99 | 0.95 | 0.98 | 0.99 |
| BRSET Image Quality Checker | 0.81 | 0.73 | 0.89 | 0.98 | 0.90 | 0.84 | 0.98 |

***Table 5:** Performance of the image quality assessment model on mBRSET and BRSET test set. The quality checker achieves reasonable to strong overall discrimination, with high AUPRC and PPV across both datasets, although specificity is lower for mBRSET.*

As seen in Table 5, the RETFoundGreen-initialized image quality model achieved reasonable performance, suggesting that the limited improvement from cascade filtering was not simply due to a poor quality classifier. Instead, the results suggest that explicit image quality labels may be weakly aligned with downstream diabetic retinopathy diagnostic performance.

To further evaluate this, we next assess whether image quality labels are associated with differences in diagnostic performance in practice (Table 6a and 6b).

| Image Quality | Number of images | Sensitivity | Specificity | Balanced Accuracy |
|---|---|---|---|---|
| Low quality | 45 | 66.7% | 87.9% | 77.3% |
| High quality | 1439 | 65.0% | 92.0% | 78.5% |

| Image Quality | Number of images | Sensitivity | Specificity | Balanced Accuracy |
|---|---|---|---|---|
| Low quality | 549 | 78.5% | 96.4% | 87.5% |
| High quality | 4011 | 80.7% | 94.4% | 87.6% |

***Table 6 (a, b):** Diagnostic performance of the single-image baseline model on the combined validation and test sets stratified by image quality on mBRSET and BRSET respectively. Comparable balanced accuracy between low- and high-quality images is observed across datasets, suggesting limited performance degradation in images labeled as poor quality.*

To further analyze model predictions across images of varying quality, we perform occlusion-based saliency analysis (Figure 5). A representative example from mBRSET is shown. This image contains strong illumination artifacts in the lower region and is therefore labeled as “poor quality.” However, the model’s saliency map highlights clinically relevant retinal structures associated with diabetic retinopathy, indicating that meaningful diagnostic features remain visible despite the artifacts. Additional saliency examples across multiple artifacts are provided in the Supplementary Materials.

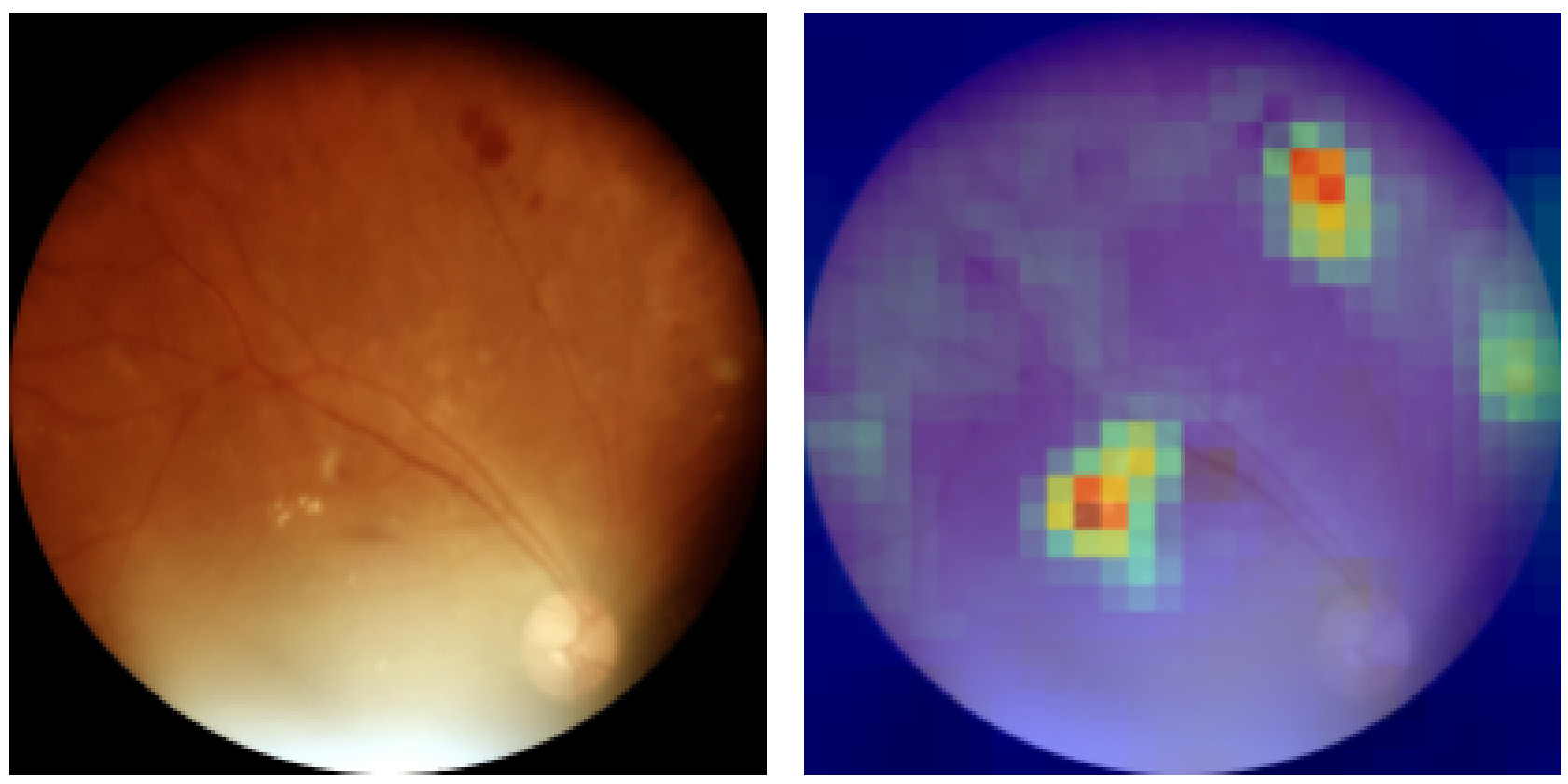

*Figure 5: Occlusion Map Based Saliency Analysis. mage quality: poor, Model Confidence: 0.99, Label: Diabetic Retinopathy*

As we adopt a confidence-based filtering approach, we explore the question of whether the model's predicted probabilities are actually correlated with its likelihood of being correct. To evaluate this, we perform a calibration analysis as shown in Figure 6 by grouping test images into confidence bins based on their predicted DR probability. For each bin, we compute the observed fraction of DR-positive cases and compare it to the mean predicted probability. Ideally, these values should match, indicating that the model's confidence accurately reflects the true probability of disease.

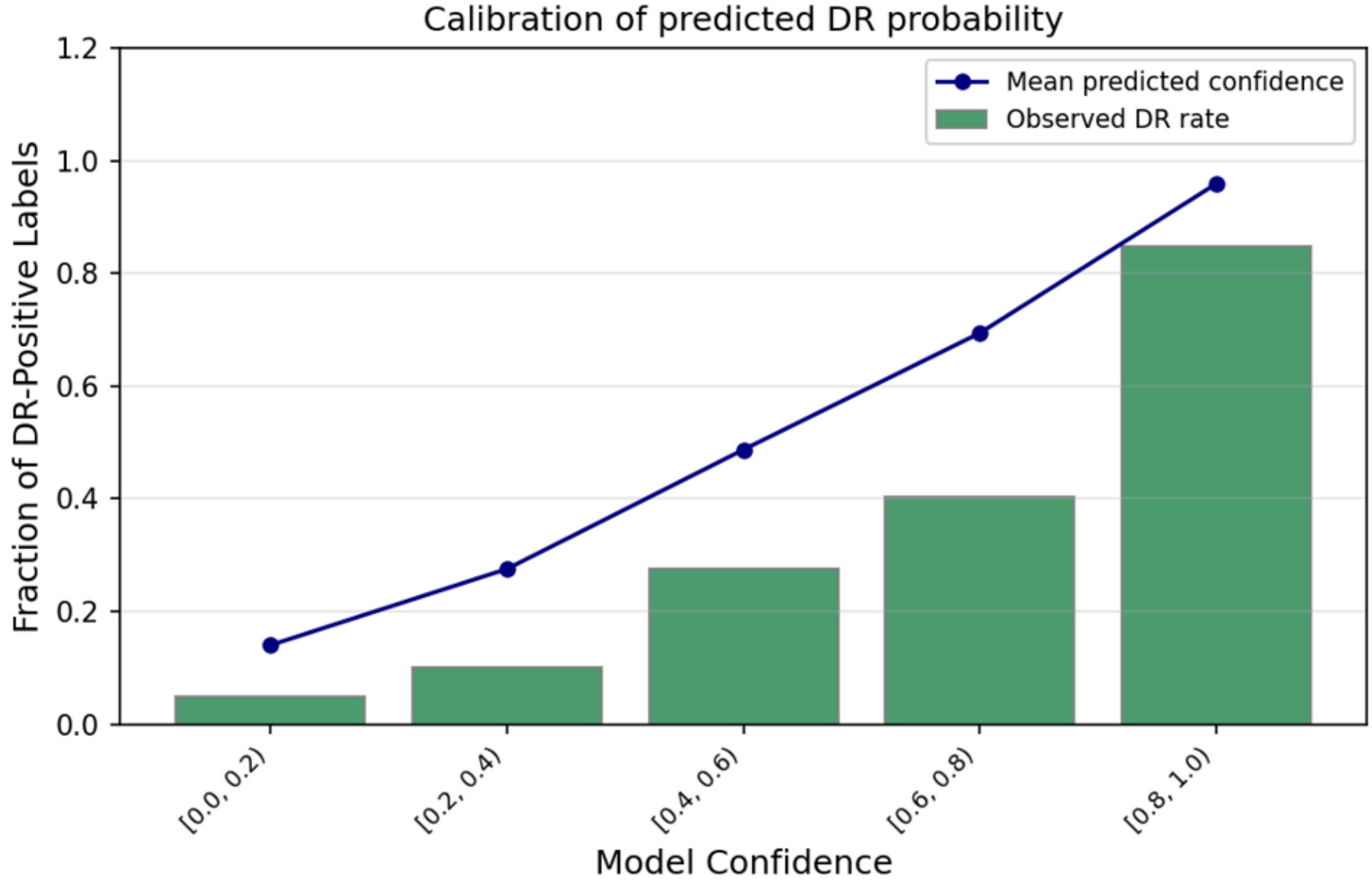

***Figure 6:*** *Calibration of predicted diabetic retinopathy probability on the validation and test sets. Green bars show the observed fraction of DR-positive cases within each predicted probability bin, while the blue line shows the mean predicted probability for that bin.*

As model confidence increases, the observed fraction of DR-positive cases also increases monotonically, suggesting that the model is directionally well-calibrated: higher confidence predictions consistently correspond to a greater prevalence of DR. However, the model exhibits slight overconfidence across most bins, as the mean predicted probability generally exceeds the observed positive rate. This indicates that the model's predicted probabilities tend to be somewhat higher than the true underlying disease prevalence.

## Discussion

In this study, we evaluated a confidence-guided multi-image fusion approach for diabetic retinopathy diagnosis and analyzed how image-level filtering and patient-level aggregation affect diagnostic reliability.

A large body of prior work has focused on fundus image quality assessment, developing increasingly sophisticated methods to classify images as gradable or ungradable based on visual features such as illumination, contrast, and clarity. One such work is DeepFundus[25], which uses a large Inception-ResNet-V2 backbone and a collection of 12 separate classifiers to assign detailed quality labels, such as illumination, positioning, clarity, and visibility of specific retinal structures like the macula or optic disc. This makes the system highly sophisticated and computationally intensive, and it performs well at identifying a wide range of quality defects on large datasets. However, when used as a preprocessing step for downstream diabetic retinopathy detection, the resulting improvement in diagnostic accuracy is relatively small, on the order of only a few percentage points. Consistent with these findings, we observe that quality-based cascade filtering provides only limited gains and quickly plateaus as coverage increases, indicating that excluding images based on predefined quality criteria does not meaningfully improve classification reliability. We explored this in more detail through occlusion map-based saliency analysis and found that conventional image quality labels do not necessarily reflect the true diagnostic utility of a fundus image, as images graded as low quality may still contain sufficient information for accurate model predictions.

Our calibration analysis supports the use of prediction confidence for selective deferral. Specifically, higher predicted DR probabilities correspond to higher observed DR prevalence, although predicted probabilities tend to exceed observed event rates across most bins. Thus, the model is directionally informative but somewhat overconfident, suggesting that confidence-based deferral is useful for ranking prediction reliability, while probability values should not be interpreted as perfectly calibrated risk estimates. In a screening context, modest overestimation of DR probability may be acceptable, and potentially preferable to underestimation, when operating thresholds are selected to prioritize sensitivity and minimize missed disease. However, calibration remains important, particularly for high-confidence negative predictions, as overconfident negatives could lead to inappropriate reassurance.

Alternative strategies have explored uncertainty-aware prediction and rejection mechanisms to improve diagnostic performance.[26-27] These approaches typically estimate predictive uncertainty using probabilistic or Bayesian deep learning methods, which approximate a distribution over predictions (mean and variance) and allow systems to defer decisions when uncertainty is high. While such strategies can reduce diagnostic risk in diabetic retinopathy screening, they often require repeated sampling and computationally intensive probabilistic modeling (10-30x inferences per image). In contrast, our confidence-based approach requires only one inference pass per image and shows significant improvements compared to an image-quality-based cascade approach.

In mBRSET, each patient is represented by multiple fundus images, including left- and right-eye images as well as repeated captures from the same view. This makes patient-level fusion different from prior multi-view fundus approaches that primarily combine anatomically distinct fields. In our setting, both eyes may provide clinically relevant information because diabetic retinopathy can present asymmetrically, while repeated same-view images may differ in quality, focus, illumination, artifacts, or lesion visibility. Fixed aggregation methods such as probability averaging or max pooling[28] may therefore be limited: averaging treats all images as equally informative, while max pooling can overemphasize a single high-confidence prediction and is mainly optimized for sensitivity. We instead use transformer-based fusion to aggregate image-level embeddings across the full patient image set, allowing the model to learn which images and features to emphasize when making the final diagnosis.

Prior work has also approached multi-image modeling using more structured and constrained architectures such as DMS-Net and binocular CNN frameworks. DMS-Net[14] proposes a binocular fusion framework that processes two fundus photographs with a common backbone. However, this approach relies on a complex fusion algorithm with several nested components, making the overall system more complicated and harder to deploy. Related binocular CNN approaches similarly take left- and right-eye fundus images as paired inputs and produce separate eye-level predictions; however, these methods are constrained to exactly one image per eye and do not aggregate information at the patient level, limiting their flexibility in real-world screening settings where the number of available images per patient varies[29]. In contrast, we implement a two-layer transformer-based aggregation module that supports variable image inputs through masking and reduces complexity.

From a deployment perspective, our approach offers important advantages over prior methods. Uncertainty-aware techniques introduce significant computational overhead, limiting their applicability in low-resource settings. In contrast, our pipeline remains lightweight: the full model comprises only ~25M parameters, with just ~3.6M in the multi-image aggregation transformer. It requires only a single inference per image (~100 milliseconds on a mobile device), and can operate fully offline, making it well suited for real-time screening applications in resource-limited environments.

*Limitations*

This study has several limitations. First, both BRSET and mBRSET were collected in Brazil, which may limit the generalizability of our findings to other populations and clinical settings. Second, the datasets, particularly BRSET, exhibit class imbalance. These distributions may not fully reflect the prevalence or severity distribution of diabetic retinopathy in screening settings: real-world class imbalance may vary across populations and could be more or less extreme than observed in these datasets. Finally, our experiments focus exclusively on diabetic retinopathy prediction. In future work, we hope to evaluate this fusion framework on other ophthalmic diseases and imaging tasks where datasets with multiple fundus images per patient are available.

*Conclusions*

Overall, our findings suggest that confidence-guided multi-image fusion can improve the accuracy-coverage tradeoff for patient-level diabetic retinopathy screening, particularly when multiple fundus images are available. By using model confidence rather than predefined image-quality labels to guide deferral, the proposed framework may better preserve diagnostically useful images while reducing unreliable predictions. Moreover, this lightweight, offline-compatible approach may support more robust and accessible screening in resource-limited settings.

## Data Availability

The BRSET and mBRSET datasets used in this study are publicly available via the PhysioNet repository. BRSET (tabletop clinic-captured fundus images with metadata) can be accessed at https://physionet.org/content/brazilian-ophthalmological/1.0.1/, and mBRSET (smartphone-captured fundus images with metadata) is available at https://physionet.org/content/mbrset/1.0/.

## Code Availability

All code is open source and is available at https://github.com/Anisha234/SureSight.

## Acknowledgements

We would like to thank Professor Marina Sirota for her support in acquiring the datasets for this work.

## Author Contribution Statement

AnaR and AniR were equally responsible for developing the patient-level confidence-thresholding framework and conducting all training and testing experiments. YMP, AST, TTO, and TNK reviewed the manuscript and gave overall guidance on the ophthalmological aspects of the work. Correspondence should be directed to TTO and TNK.

Disclosure: Ananya Raghu, None; Anisha Raghu, None; A.S. Tang, None; Y.M. Paulus, None; T.N. Kim, None; T.T. Oskotsky, None

## Supplementary Information

| ***Preprocessing Step*** | ***Number of images*** |
|---|---|
| *Original dataset size* | *5,164* |
| *Fill missing ICDR labels using the paired image from the same eye* | *5,164* |
| *Remove any NaN values* | *5,046* |
| *Keep only patients with exactly two images for each eye* | *4,936 / 1,234 patients* |

***Table S1:*** *Data preprocessing steps for mBRSET dataset*

| ***Preprocessing Step*** | ***Number of images*** |
|---|---|
| *Original dataset size* | *16,266* |
| *Remove any NaN values* | *16,266* |
| *Keep only patients with exactly one image per eye* | *15,198 / 7,599 patients* |

***Table S2:*** *Data preprocessing steps for BRSET dataset*

| ***Train-Val-Test Split*** | **Number of Patients** |
|---|---|
| *Train set* | 863 |
| *Validation set* | 185 |
| *Test set* | 186 |

***Table S3:*** *Train, Val, Test Split for the mBRSET dataset, using a 70%, 15%, 15% split at the patient level*

| ***Train-Val-Test Split*** | **Number of Patients** |
|---|---|
| *Train set* | 5319 |
| *Validation set* | 1140 |
| *Test set* | 1140 |

***Table S4:*** *Train, Val, Test Split for the BRSET dataset, using a 70%, 15%, 15% split at the patient level*

### ***Saliency Analysis***

*ARTIFACT #1: Illumination*

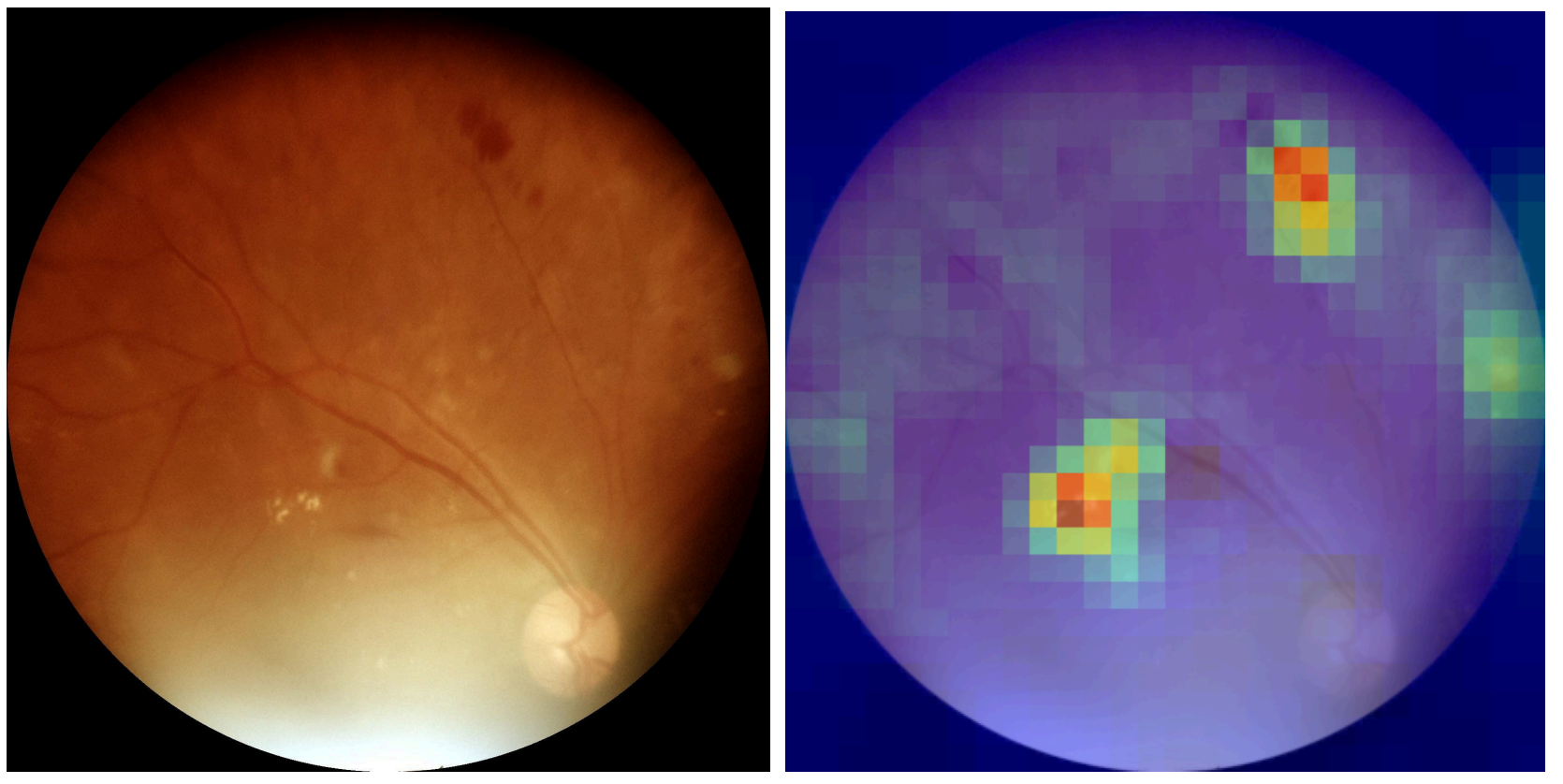

*Image quality: poor, Model Confidence: 0.99, Label: Diabetic Retinopathy*

The model is focusing on:

- Hard exudates
- Microaneurysms

*ARTIFACT #2: Occlusion*

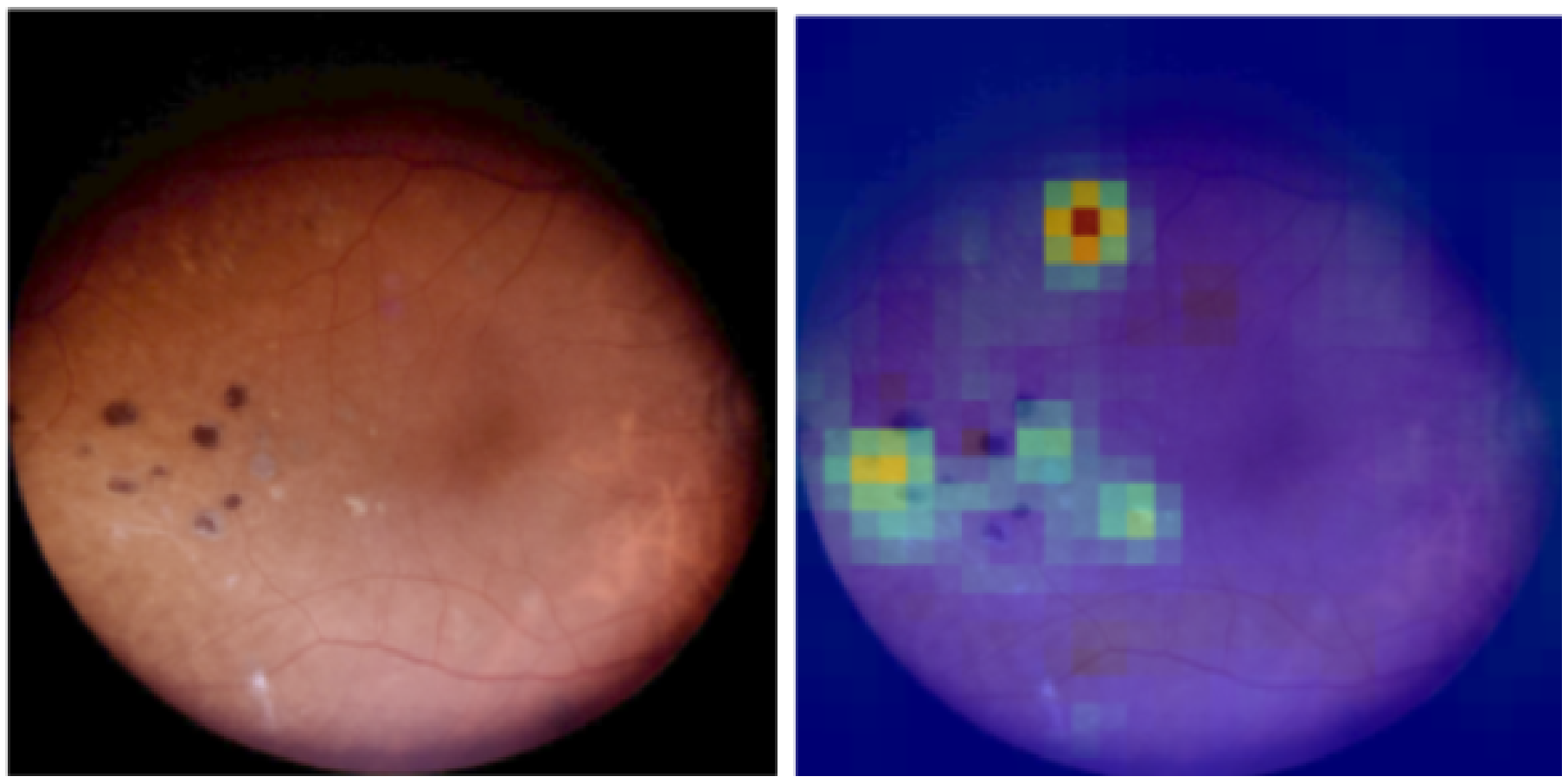

*Image quality: poor, Model Confidence: 0.99, Label: Diabetic Retinopathy*

The model is focusing on:

- Microaneurysms
- Hemorrhages

*ARTIFACT #3: Blurring*

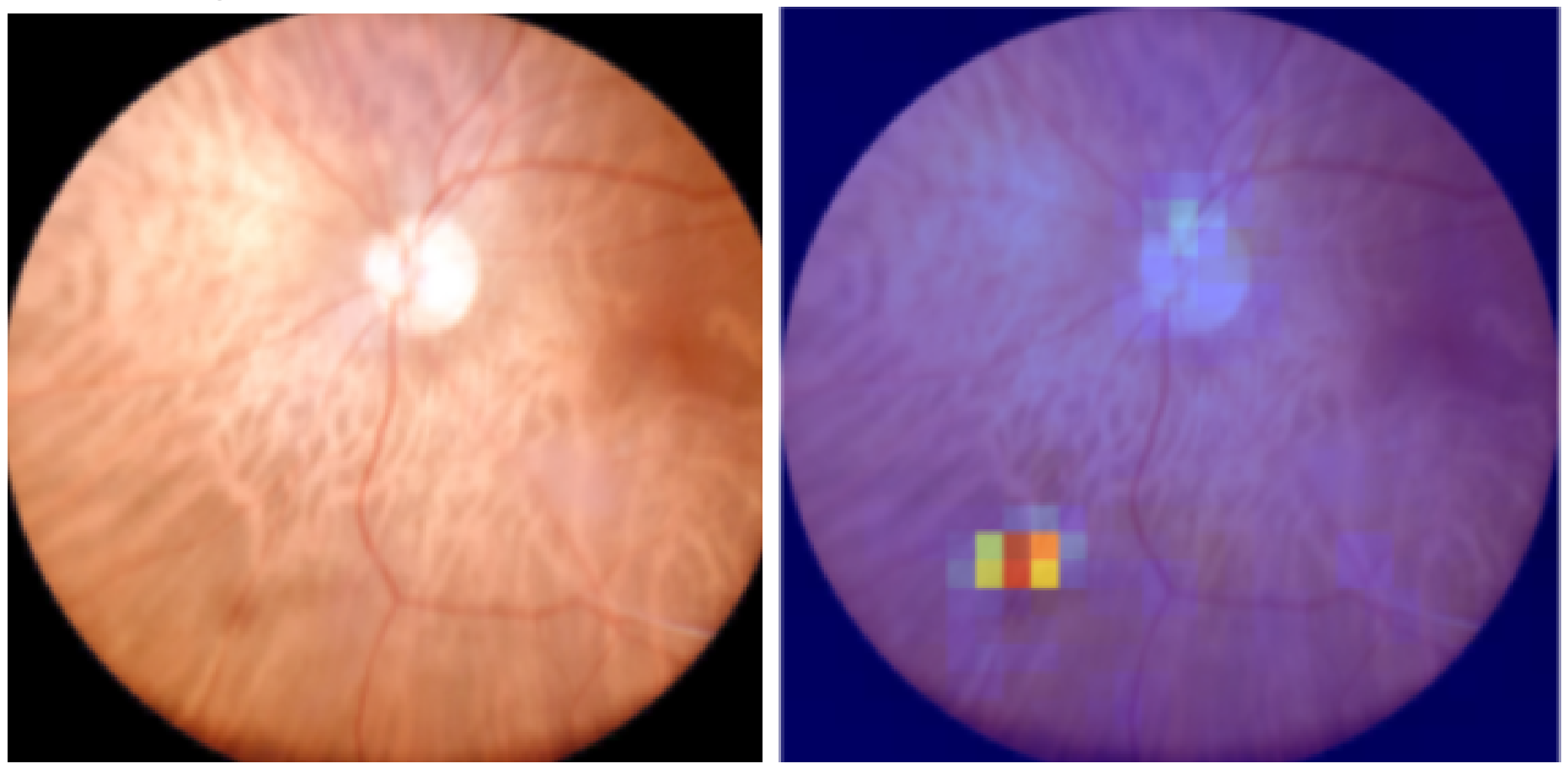

*Image quality: poor, Model Confidence: 0.90, Label: Diabetic Retinopathy*

The model is focusing on:

- Hemorrhages